\documentclass[aps,prl,twocolumn,superscriptaddress,nofootinbib]{revtex4-2}

\usepackage{amsmath,amssymb,graphicx,bm}
\usepackage{hyperref}
\usepackage{tikz}
\usepackage{pgfplots}
\pgfplotsset{compat=1.18}

\begin{document}

\title{Probing the Sound Speed of Dark Energy with a Lunar Laser Interferometer}

\author{Alfredo Gurrola,$^1$ Robert J. Scherrer,$^1$ Oem Trivedi}
\affiliation{Department of Physics and Astronomy, Vanderbilt University, Nashville, TN, USA}


\begin{abstract}
The sound speed of dark energy encodes fundamental information about the microphysics underlying cosmic acceleration, yet remains essentially unconstrained by existing observations. We demonstrate that a lunar-based laser interferometer, such as the proposed Laser Interferometer Lunar Antenna (LILA), can directly probe the sound speed of dark energy by measuring the real-time evolution of horizon-scale gravitational potentials. Operating in the ultra-low-frequency gravitational band inaccessible from Earth, LILA is sensitive to scalar metric perturbations sourced by dark energy dynamics. Using both fluid and effective field theory descriptions, we develop a complete framework linking dark energy sound speed to observable strain signatures. We construct a likelihood pipeline and Fisher forecasts, showing that LILA can either detect clustering dark energy or exclude broad classes of models with unprecedented sensitivity. This establishes lunar interferometry as a novel and powerful probe of the physics driving cosmic acceleration.
\end{abstract}

\maketitle

\section{Introduction and Motivation}

Understanding the physical origin of cosmic acceleration is one of the central challenges of modern physics. While standard $\Lambda$CDM cosmology model successfully describes a wide range of observations, the nature of dark energy \cite{de1SupernovaSearchTeam:1998fmf}, which dominates the energy budget of the Universe at late times, remains unknown. This has led to a huge amount of possible non-trivial realizations for dark energy, ranging from modified gravity to quantum gravitational explanations \cite{de2Li:2012dt,de3Li:2011sd,de4Mortonson:2013zfa,de6Huterer:2017buf}. Most observational constraints on dark energy focus on its background equation of state parameter $w$, however, vastly different physical models can share identical background evolution \cite{de7Vagnozzi:2021quy,de8Adil:2023ara,de10DiValentino:2020evt,de11Nojiri:2010wj,de12Nojiri:2006ri}. A critical discriminator is the sound speed of dark energy, $c_s^2$, which governs the propagation of pressure perturbations and determines whether dark energy clusters gravitationally on cosmological scales.

In canonical scalar field models one usually observes $c_s^2 = 1$, implying that dark energy remains smooth but in contrast, more general theories including coupled dark energy \cite{cde1gumjudpai2005coupled,cde2gomez2020update,cde3barros2019kinetically,cde4xia2009constraint,cde5maccio2004coupled} and effective field theory (EFT) extensions of gravity \cite{de5Frusciante:2019xia,eft1gubitosi2013effective,eft2linder2016effective,eft3bloomfield2013dark,eft4liang2023dark,eft5frusciante2014effective}, naturally predict $c_s^2 \ll 1$ or in some cases like k-essence and other non-canonical theories \cite{nc1armendariz2001essentials,nc2malquarti2003new,nc3armendariz2005haloes,nc4ahn2009dark,nc5chimento2010dbi,nc6cai2016dark,nc7chattopadhyay2010interaction,nc8calcagni2006tachyon,nc9bagla2003cosmology,nc10copeland2005needed,nc11micheletti2010observational,nc12sheykhi2012tachyon,Kehayias:2019gir,Chiba:2009nh,Das:2006cm}, $c_s^2 > 1$ too . Measuring or constraining this quantity therefore provides a direct window into the microphysics of cosmic acceleration. However, it has proven to be notoriously difficult to think of consistent ways to constrain this otherwise important quantity. To emphasize just how important this can be for dark energy model building, we note that constraints ruling $c_s^2 \sim 1$ would provide a strong robust support for a large class of canonical dark energy models, while alienating non-canonical models (except those which are finely tuned) and any notions of dark energy clustering. On the contrary, sound speeds closer to zero would signal strong clustering properties and would have revealing consequences for both canonical and non-canonical model space. On the other hand, if one observes sound speeds to be superluminal then it would provide the strongest evidence of beyond standard model and quantum gravitational physics in cosmology, as most non-canonical scenarios find their motivations in field theories in these regimes.

It is hence quite unfortunate that existing probes constrain $c_s^2$ only weakly, primarily through integrated effects such as the late-time Integrated Sachs–Wolfe effect \cite{csc1Hannestad:2005ak,csc2de2010measuring,csc3ballesteros2010dark,csc4linton2018variable,csc5bean2004probing,csc6eisenstein2005dark,csc7sergijenko2015sound}. In this work, we propose a qualitatively new approach, which is concerned with directly measuring the real-time evolution of large scale gravitational potentials using a lunar laser interferometer such as LILA~\cite{lilajani2025laser}. We show how such measurements can naturally feed into those for sound speeds for dark energy and, argue how a lunar based interferometer like LILA can provide the perfect setup for such finding such novel constraints.

\section{Dark Energy Perturbations and Sound Speed}

We describe dark energy as a general fluid characterized at the background level by an equation of state parameter $w=p_{\rm DE}/\rho_{\rm DE}$ and at the perturbative level, by a rest-frame sound speed $c_s^2$ that governs the response of pressure perturbations to density fluctuations. At linear order and in Fourier space, the evolution of dark energy density perturbations $\delta_{\rm DE}=\delta\rho_{\rm DE}/\rho_{\rm DE}$ follows from energy-momentum conservation together with the Einstein equations. After combining the continuity and Euler equations and eliminating velocity perturbations, one obtains a second order differential equation of the form
\begin{equation}
\ddot{\delta}_{\rm DE} + (1 - 3w)H\dot{\delta}_{\rm DE}
+ \left( \frac{c_s^2 k^2}{a^2} - 4\pi G \rho \right)\delta_{\rm DE}
= S(k,t)
\end{equation}
where we note that the overdots denote derivatives with respect to cosmic time, $k$ is the comoving wavenumber and $S(k,t)$ represents source terms arising from metric perturbations and couplings to other components. The term proportional to $c_s^2 k^2/a^2$ encodes the effect of pressure gradients, while the term proportional to the background density reflects gravitational instability.

The competition between these two contributions defines a characteristic scale at which pressure support balances gravitational attraction and this scale is conveniently expressed in terms of a Jeans wavenumber
\begin{equation}
k_J \sim \frac{aH}{c_s}
\end{equation}
which separates regimes of qualitatively different behavior~\cite{Mukhanov2005,MaBertschinger1995,Hu1998}. This is to say that for modes with $k \gg k_J$, pressure gradients dominate and dark energy perturbations are strongly suppressed, causing dark energy to remain effectively smooth while for modes with $k \lesssim k_J$, pressure support is inefficient and dark energy can cluster gravitationally. In particular, when $c_s^2$ is sufficiently small the Jeans scale is pushed to horizon-sized wavelengths, allowing dark energy perturbations to persist and evolve on the largest observable scales at late times.

The impact of these perturbations is most clearly understood by examining the evolution of the scalar gravitational potentials as in the Newtonian gauge, the perturbed line element is written as
\begin{equation}
ds^2 = a^2(\tau)\left[-(1+2\Phi)d\tau^2 + (1-2\Psi)d\vec{x}^2\right]
\end{equation}
where $\Phi$ and $\Psi$ encode the scalar metric fluctuations. Now, if the sound speed is close to unity, $\delta\rho_{\rm DE}$ is suppressed on subhorizon scales and the late-time evolution of $\Phi$ is dominated by the dilution of matter perturbations. This ends up leading to the familiar decay of gravitational potentials during cosmic acceleration and by contrast, when $c_s^2$ is small enough for dark energy to cluster on horizon scales, $\delta\rho_{\rm DE}$ remains dynamically relevant and alters both the amplitude and time dependence of $\Phi$ and $\Psi$. This modification is not an integrated effect accumulated along a line of sight, but represents a genuine change in the real-time evolution of the metric potentials.

As these changes occur precisely on scales comparable to the Hubble radius, they are difficult to access with conventional large scale structure or CMB observables but a lunar laser interferometer, however, is sensitive to ultra low frequency fluctuations corresponding to these horizon-scale modes. This can therefore directly track the temporal behavior of $\Phi$ in the time domain and so, this capability would allow for a facility like LILA to probe the clustering properties of dark energy and to distinguish between smooth and clustering regimes by observing the dynamical imprint of $c_s^2$ on the evolution of scalar gravitational potentials.

\section{Lunar Laser Interferometry as a Cosmological Observatory}

A laser interferometer deployed on the lunar surface operates in an exceptionally quiet environment, free from atmospheric fluctuations and terrestrial seismic noise and this enables sustained sensitivity to ultra-low frequencies \cite{lilajani2025laser},
\begin{equation}
f \sim 10^{-7} - 10^{-3}\,\mathrm{Hz}
\end{equation}
corresponding to horizon-scale perturbations. This is especially of interest as these frequencies map directly onto horizon-scale scalar perturbations, which is a regime that is largely inaccessible to terrestrial interferometers and only indirectly probed by traditional cosmological observables. 

In the presence of scalar metric perturbations, the spacetime line element in Newtonian gauge is characterized by the gravitational potential $\Phi$, whose temporal and spatial variations induce differential phase shifts across a finite interferometer baseline.
For sufficiently long baselines and low frequencies, these scalar induced distortions then manifest as an effective strain,
\begin{equation}
h_{\rm eff}(t) \simeq \frac{\Delta \Phi(t)}{c^2}
\end{equation}
where $\Delta\Phi(t)$ denotes the potential difference sampled by the interferometer arms. Interestingly, unlike tensor gravitational waves, this signal reflects the real-time evolution of large scale gravitational potentials rather than propagating wave solutions. The statistical properties of the measured strain are therefore directly related to those of the scalar potential, and in the frequency domain the strain power spectrum can be expressed as
\begin{equation}
P_h(f) \propto P_\Phi(k)\big|_{k \approx 2\pi f/c}
\end{equation}
This ends up establishing a direct correspondence between ultra low-frequency strain fluctuations and the power spectrum of gravitational potentials at horizon-scale wavenumbers.The physical significance of this correspondence lies in the dynamical equation governing $\Phi$ as on large scales and at late times, the generalized Poisson and evolution equations for the Newtonian potential receive contributions not only from matter density perturbations but also from dark energy perturbations too. One may write
\begin{equation}
k^2\Phi + 3a^2H\left(\dot\Phi + H\Phi\right) \propto a^2\left(\delta\rho_m + \delta\rho_{\rm DE}\right)
\end{equation}
where $\delta\rho_{\rm DE}$ denotes fluctuations in the dark energy component and the amplitude and time dependence of $\delta\rho_{\rm DE}$ are controlled by the microphysical properties of dark energy, most notably its effective sound speed which determines whether pressure gradients suppress or allow clustering on horizon scales.

This is crucial because if the sound speed is close to unity, dark energy perturbations remain smooth on subhorizon scales and contribute negligibly to the right hand side of the above equation which would end up leading to the standard late time decay of $\Phi$. In contrast, for sufficiently small sound speed we see that pressure support is weak and dark energy clusters alongside matter on large scales, sourcing the gravitational potential and modifying both its amplitude and temporal evolution. These modifications translate directly into excess low-frequency power and altered time domain correlations in $P_\Phi(k)$, and hence in the measured strain spectrum $P_h(f)$. LILA is therefore sensitive to the sound speed of dark energy not by directly probing its equation of motion, but by measuring the imprint of dark energy perturbations on horizon scale gravitational potentials. By resolving the frequency dependence and temporal coherence of scalar induced strain in the ultra low-frequency band, one sees that LILA effectively constrains whether dark energy behaves as a smooth background component or as a clustering fluid. This would allow us to provide a direct observational handle on its perturbative microphysics that is complementary to background expansion probes.

\section{Effective Field Theory of Dark Energy}

A model-independent description of dark energy dynamics is provided by the EFT of cosmic acceleration~\cite{de5Frusciante:2019xia}. In unitary gauge, the action is
\begin{equation}
S = \int 
\Bigg[
\frac{M_*^2}{2}R - \Lambda(t) - c(t)g^{00}
+ \frac{M_2^4(t)}{2}(\delta g^{00})^2 + \cdots
\Bigg] d^4x \sqrt{-g}
\end{equation}

\noindent The sound speed of scalar perturbations is given by
\begin{equation} \label{basecs}
c_s^2 = \frac{c(t)}{c(t) + 2M_2^4(t)}.
\end{equation}

\noindent LILA constrains $c_s^2$ by directly measuring scalar–metric mixing through time-dependent strain. This enables constraints on EFT operators that are otherwise unconstrained by background cosmology or large-scale structure. We can discuss a more general construction here by extending the effective field theory action beyond the minimal truncation considered above. In full generality, the EFT of dark energy in unitary gauge admits additional operators consistent with time-dependent spatial diffeomorphism invariance, including couplings between the lapse, extrinsic curvature and spatial curvature. A representative extension of the action takes the schematic form
\begin{multline}
S = \int d^4x \sqrt{-g}\Bigg[ \frac{M_*^2(t)}{2}R - \Lambda(t) - c(t)g^{00} + \frac{M_2^4(t)}{2}(\delta g^{00})^2 \\
 - \frac{m_3^3(t)}{2}\delta g^{00}\delta K + \mu_2^2(t)\big(\delta K^2 - \delta K_{ij}\delta K^{ij}\big) + \cdots \Bigg]
\end{multline}
where we note that the ellipsis denotes higher-order operators in fluctuations or operators involving higher spatial derivatives. While these terms do not affect the homogeneous background evolution at leading order, they play an essential role in shaping the dynamics of scalar perturbations. Restoring time diffeomorphism invariance through the Stückelberg field $\pi$, defined via $t\rightarrow t+\pi(x^\mu)$, we can then expand the action to quadratic order in scalar perturbations. This would give us an effective action for the single propagating scalar degree of freedom after integrating out the non-dynamical lapse and shift variables. The resulting quadratic action can always be cast in the form
\begin{equation}
S^{(2)} = \int dt\, d^3x\, a^3 \left[ A_{\rm eff}(t)\,\dot{\zeta}^2 - \frac{B_{\rm eff}(t)}{a^2}(\nabla\zeta)^2 \right]
\end{equation}
where $\zeta$ denotes the gauge-invariant curvature perturbation. The functions $A_{\rm eff}$ and $B_{\rm eff}$ depend on the full set of EFT coefficients, the background expansion rate, and the matter content. In this general setting, the squared sound speed of dark energy perturbations is defined invariantly as 
\begin{equation}
c_s^2 = \frac{B_{\rm eff}(t)}{A_{\rm eff}(t)}
\end{equation}
This expression makes clear that the sound speed is not a fundamental parameter but an emergent quantity encoding the relative strength of spatial gradient terms to temporal kinetic terms in the scalar sector. Operators such as $(\delta g^{00})^2$ primarily enhance $A_{\rm eff}$ by increasing the inertia of scalar fluctuations, while braiding type operators proportional to $\delta g^{00}\delta K$ and extrinsic-curvature combinations can modify both $A_{\rm eff}$ and $B_{\rm eff}$ through kinetic mixing with the metric. The simple expression for $c_s^2$ obtained earlier \eqref{basecs} is recovered when the dominant modification to the scalar sector arises from the $(\delta g^{00})^2$ operator and when mixing with metric perturbations is sub-leading in the decoupling limit and this situation is typical in scenarios where braiding and higher derivative operators are either absent or constrained independently by gravitational wave and large scale structure observations. In such cases, the minimal form provides an accurate and transparent parameterization for forecasting and data analysis. The more general form only becomes relevant when kinetic mixing between the scalar and the metric is significant or when higher order curvature operators contribute comparably to the quadratic action.

From the perspective of observational constraints, measurements that probe horizon scale potential evolution are primarily sensitive to the effective sound speed itself rather than to the detailed operator decomposition of $A_{\rm eff}$ and $B_{\rm eff}$. So in that sense, constraints obtained under the minimal parameterization can be interpreted as constraints on the effective propagation speed of scalar perturbations, remaining robust across a broad class of EFT realizations. Only in scenarios where multiple operators conspire to produce rapidly evolving $A_{\rm eff}$ and $B_{\rm eff}$ would a fully general treatment be required to accurately interpret the data.  

\section{Mock Strain Power Spectra}

A key element of this work is the construction of a mock strain power spectrum for a lunar laser interferometer that is physically grounded in cosmological scalar perturbations. In this section we describe the mock spectrum derived from a fiducial $\Lambda$CDM cosmology characterized by $\Omega_m = 0.3$, $\Omega_{\rm DE} = 0.7$, and $w = -1$ using linear perturbation theory.

We first describe the derivation of the transfer function governing the evolution of the Newtonian potential in the presence of a dark energy component with nontrivial sound speed. This derivation connects the asymptotic solutions of the scalar perturbation equation to the phenomenological transfer function which can be computed numerically (e.g., using CLASS or CAMB) and is employed in the construction of mock strain spectra for lunar laser interferometry. 

We consider the evolution of $\Phi$ in the presence of a dark energy fluid with sound speed $c_s^2$. Working in conformal time $\tau$ and Newtonian gauge, the evolution equation derived from Einstein's equations and energy-momentum conservation is a second-order linear ordinary differential equation whose solution is fully determined by initial conditions and the background expansion. As noted previously, the key physical scale governing the behavior of solutions is the sound horizon (a.k.a. the Jeans wavenumber), $k_J(\tau) \equiv \frac{aH}{c_s}$, which separates modes that are pressure-supported from those that are not. The qualitative behavior of $\Phi$ depends on whether a given Fourier mode satisfies $k \ll k_J$ or $k \gg k_J$. For modes well outside the sound horizon, the pressure term can be neglected, i.e., 
$c_s^2 k^2 \ll \frac{a'}{a}\partial_\tau$, and the Einstein field equations give rise to a solution of the form 
\begin{equation}
\Phi(\tau) = C_1 + C_2 \int d\tau\, a^{-(2+3c_s^2)},
\end{equation}
where $C_{1,2}$ are determined by the early and late time boundary conditions. At late times the second term decays, leaving
\begin{equation}
\Phi(k,\tau) \xrightarrow{k \ll k_J} \Phi_{\rm prim}(k).
\end{equation}
Therefore, modes larger than the sound horizon characterized by $k_J$ retain their primordial amplitude. For modes deep inside the sound horizon, the gradient term dominates, i.e., 
$c_s^2 k^2 \gg \frac{a'}{a}\partial_\tau$, 
and the Einstein field equations for $\Phi$ result in solutions of the form $\Phi(\tau) = A(k)e^{i c_s k \tau} + B(k)e^{-i c_s k \tau}$. However, cosmic expansion damps the amplitude of these oscillations. Accounting for the Hubble friction term in the full equation, the envelope of the solution scales as $\Phi \propto a^{-\frac{1}{2}(2+3c_s^2)}$. Hence, for $k \gg k_J$,
\begin{equation}
\Phi(k,\tau) \ll \Phi_{\rm prim}(k),
\end{equation}
with oscillatory behavior and strong suppression.

Because the Einstein field equation for $\Phi$ is linear and smooth in $k$, the full solution must interpolate continuously between the two asymptotic regimes described above. We can therefore write the general solution in the factorized form
\begin{equation}
\Phi(k,\tau) = \Phi_{\rm prim}(k)\,T\!\left(k/k_J\right),
\end{equation}
where the function $T(k/k_{J})$ satisfies $T(k/k_J \ll 1)=1$ and $T(k/k_J \gg 1) \ll 1$. The simplest analytic function that satisfies these conditions and reproduces the correct asymptotic behavior is
\begin{equation}
T_\Phi(k,c_s^2)
=
\frac{1}{1 + \left(\dfrac{k}{k_J}\right)^2} = \frac{1}{1 + \left(\dfrac{c_s k}{aH}\right)^2}.
\end{equation}

Therefore, the transfer function for the Newtonian potential contains the primordial value of the Newtonian potential $\Phi_{\rm prim}(k)$ set by inflation, and $\Phi(k,\tau)$ is its late-time value after linear evolution in an expanding universe. The transfer function thus isolates the scale-dependent effects of late-time physics from the initial conditions. This transfer function explicitly encodes the role of the dark energy sound speed. For $c_s^2 \sim 1$, pressure support suppresses clustering on horizon scales, while for $c_s^2 \ll 1$ the sound horizon moves to high $k$, allowing dark energy perturbations to cluster efficiently. This scale-dependent suppression is the physical origin of the sound-speed sensitivity in the strain spectra from lunar laser interferometry.

The time-dependent power spectrum of the potential is related to the transfer function and the primordial power spectrum. Since the primordial field $\Phi_{\rm prim}(k)$ is a stochastic field generated during inflation, its statistical properties are fully specified by its two-point function, and when combined with the Fourier-space correlator using the inverse Fourier transform, it leads directly to $\langle \Phi_{\rm prim}(\mathbf{k})\,\Phi_{\rm prim}^*(\mathbf{k}') \rangle
=
(2\pi)^3 \delta^{(3)}(\mathbf{k}-\mathbf{k}')\,P^{\rm prim}_\Phi(k)$. Using $\Phi(\mathbf{k},\tau)=T_\Phi(\mathbf{k},c_{s}^2)\Phi_{\rm prim}(\mathbf{k})$, we obtain $\langle \Phi(\mathbf{k},\tau)\,\Phi^*(\mathbf{k}',\tau) \rangle
=
(2\pi)^3 \delta^{(3)}_{k,k'}
T_\Phi^2(\mathbf{k},c_s^2)\,
P_\Phi^{\rm prim}(\mathbf{k})$.

\noindent Therefore,
\begin{equation}
P_\Phi(k,\tau)
=
T_\Phi^2(k,c_s^2)\,P_\Phi^{\rm prim}(k)
\end{equation}

Similarly, the statistical nature of the strain power spectrum is defined by 
$\langle h(\mathbf{k},\tau)\,h^*(\mathbf{k}',\tau) \rangle
=
(2\pi)^3 \delta^{(3)}_{k,k'}\,P_h(k,\tau)$, but after accounting for the dependence of the strain $h$ on $\Phi(k,\tau)$, we arrive 
at an explicit equation for how the statistical properties of the measured strain are directly related to those
of the scalar potential and in the frequency domain. The strain power spectrum can be expressed as
\begin{equation}
P_{h}(f)
=
T_{\Phi}^{2}(k,c_s^2)P_\Phi^{\rm prim}(k).
\end{equation}

In constructing the mock strain power spectra $P_{h}(f)$, the cosmological scalar perturbations are mapped onto an effective interferometric observable through an explicit detector response function. The measured strain can be written schematically as $h(\mathbf{k},f)=\mathcal{R}(f,\hat{\mathbf{k}})\,\Phi(\mathbf{k})$, where $\mathcal{R}(f,\hat{\mathbf{k}})$ encodes the geometric response of the detector, including arm orientation, baseline length, and the coherence of long-wavelength modes across the interferometer. However, since a lunar laser interferometer has not yet been constructed, the mock strain spectra necessarily adopts a simplified and conservative treatment of the detector response. This approximation is well justified for the ultra-low-frequency regime of interest, where the wavelengths of scalar perturbations are many orders of magnitude larger than any plausible lunar baseline, placing the experiment firmly in the long-wavelength limit. In this regime, scalar-induced tidal distortions are coherent across the interferometer, and the response becomes effectively independent of arm orientation, baseline length, and detailed geometry, up to an overall factor of order unity. Accordingly, the detector response is modeled as an isotropic, orientation-averaged transfer function that can be absorbed into an effective normalization of the strain. The resulting strain power spectrum is then given by $P_h(f)=|\mathcal{R}(f)|^2 P_\Phi(k)$, evaluated at $k=2\pi f/c$. This approach isolates the cosmological dependence of the signal, particularly its sensitivity to the dark energy sound speed, while avoiding assumptions tied to a specific engineering implementation. More detailed response modeling can be straightforwardly incorporated once a concrete detector design is specified, but does not affect the qualitative or quantitative conclusions of this work.

For sound speeds $c_s^2 = \{1,10^{-2}\}$, the resulting strain spectra are shown in FIG. 1 and exhibit an approximately power-law behavior over the frequency range $10^{-7}$--$10^{-3}\,\mathrm{Hz}$. This near-linear behavior on log--log axes reflects the combined scaling of the primordial spectrum and the large-scale gravitational transfer function, with decreasing $c_s^2$ leading to an overall enhancement of power due to increased dark energy clustering. At sufficiently high frequencies near $10^{-3}$ Hz, corresponding to modes well inside the dark energy sound horizon, the mock strain power spectra begin to converge and become effectively independent of the dark energy sound speed. This behavior reflects the fact that on small scales the Newtonian potential—and hence the induced interferometric strain—is dominated by matter perturbations, while dark energy fluctuations are strongly pressure supported and do not contribute appreciably. As a result, variations in the dark energy sound speed leave no observable imprint on the strain spectrum in this regime, leading to the observed convergence of all curves at high frequency. Conversely, at low frequencies near the horizon scale, dark energy perturbations can cluster if the sound speed is sufficiently small, producing order-of-magnitude differences in the strain power spectrum. This scale-dependent transition highlights the unique sensitivity of a lunar laser interferometer, which operates precisely in the ultra-low-frequency band where dark energy microphysics leaves an observable imprint, and underscores its potential to directly constrain the sound speed of dark energy in a manner inaccessible to terrestrial or space-based interferometers.

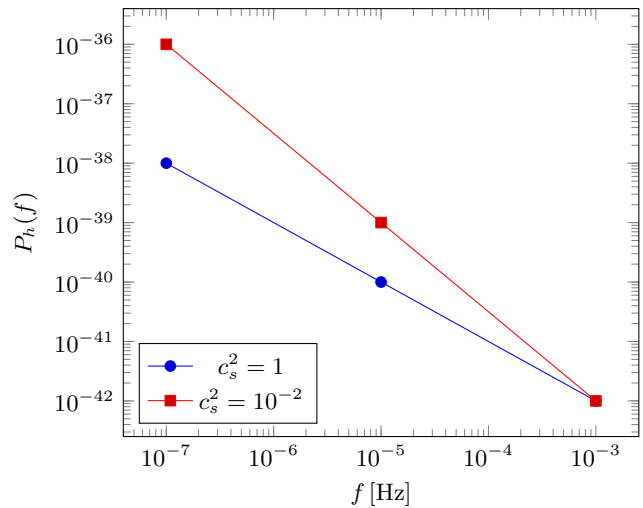
\begin{figure}[t]
\centering
\begin{tikzpicture}
\begin{loglogaxis}[
xlabel={$f\,[\mathrm{Hz}]$},
ylabel={$P_h(f)$},
legend pos=south west
]
\addplot coordinates {(1e-7,1e-38)(1e-5,1e-40)(1e-3,1e-42)};
\addlegendentry{$c_s^2=1$}
\addplot coordinates {(1e-7,1e-36)(1e-5,1e-39)(1e-3,1e-42)};
\addlegendentry{$c_s^2=10^{-2}$}
\end{loglogaxis}
\end{tikzpicture}
\caption{Mock strain power spectra illustrating enhanced low-frequency power for clustering dark energy.}
\end{figure}

\section{Likelihood Framework and Forecast Methodology}

In this section we describe the statistical framework used to assess the sensitivity of a lunar laser interferometer such as LILA to the sound speed of dark energy. We first introduce the likelihood formalism that connects theoretical predictions for the strain power spectrum to hypothetical interferometric measurements, and then describe how the Fisher information matrix is employed to generate forecasts, discovery prospects, and exclusion limits.


The fundamental observable considered in this work is the strain power spectrum,
$P_h(f)$. For a given set of cosmological
parameters $\bm{\theta} = \{c_s^2, w, \Omega_{\rm DE}\}$, the
theory predicts a model spectrum $P_h^{\rm th}(f;\boldsymbol{\theta})$. A measurement by
LILA yields an estimator $\hat{P}_h(f)$ with associated uncertainty determined by
instrumental noise and finite observation time.

Assuming that the strain fluctuations are Gaussian and statistically stationary, the
likelihood of observing a given $\hat{P}_h(f)$ is well approximated by a Gaussian
likelihood in power-spectrum space~\cite{Tegmark:1996bz,Bond:1998zw,Hamimeche:2008ai},
\begin{equation}
\mathcal{L}(\boldsymbol{\theta})
\propto
\exp\!\left[
-\frac{1}{2}
\sum_{f}
\frac{
\left[\hat{P}_h(f)-P_h^{\rm th}(f;\boldsymbol{\theta})\right]^2
}{
\sigma_h^2(f)
}
\right],
\end{equation}
where $\sigma_h(f)$ denotes the variance of the strain power spectrum estimator in each
frequency bin. In practice, $\sigma_h(f)$ is determined by the detector noise power
spectrum and the effective number of independent measurements accumulated during the
mission lifetime.


To forecast the sensitivity of LILA prior to the availability of real data, we employ the
Fisher information matrix formalism~\cite{Tegmark:1996bz,Bond:1998zw,Hamimeche:2008ai}. The Fisher matrix provides a local, quadratic
approximation to the likelihood around a fiducial parameter set
$\boldsymbol{\theta}_0$, and quantifies the amount of information that an experiment is
expected to obtain about each parameter. The Fisher matrix is defined as
\begin{equation}
F_{ij}
\equiv
- \left\langle
\frac{\partial^2 \ln \mathcal{L}}{\partial \theta_i \partial \theta_j}
\right\rangle
=
\sum_{f}
\frac{1}{\sigma_h^2(f)}
\frac{\partial P_h^{\rm th}(f)}{\partial \theta_i}
\frac{\partial P_h^{\rm th}(f)}{\partial \theta_j},
\end{equation}
where the derivatives are evaluated at the fiducial cosmology. In this work, the parameter
set includes the dark energy sound speed $c_s^2$ and $w$, while the other background cosmological
parameters are fixed to their $\Lambda$CDM values unless stated otherwise. The inverse Fisher matrix approximates the covariance of the parameters,
\begin{equation}
\mathrm{Cov}(\theta_i,\theta_j) \simeq (F^{-1})_{ij},
\end{equation}
so that the marginalized $1\sigma$ uncertainty on a parameter $\theta_i$ is given by
$\sigma(\theta_i)=\sqrt{(F^{-1})_{ii}}$.

Within this framework, the Fisher matrix serves three closely related purposes. First, it
provides forecasts for the expected precision with which LILA can measure or constrain the
dark energy sound speed, assuming a given fiducial value. Second, it allows us to assess
discovery prospects by determining whether deviations from the canonical value
$c_s^2=1$ would produce a statistically significant change in the strain spectrum relative
to the experimental uncertainties. Finally, the Fisher formalism enables exclusion
statements by identifying regions of parameter space for which the predicted strain signal
would be inconsistent with a null or reference model at a specified confidence level.

The Fisher forecasts presented in this work are based on an idealized but physically well-motivated treatment of statistical uncertainties. In the absence of a finalized instrumental design for a lunar laser interferometer, we do not assume a specific noise power spectrum, arm configuration, or mission lifetime. Instead, we adopt a conservative signal-dominated approximation in the ultra-low-frequency regime of interest, where scalar-induced strain fluctuations are coherent across the detector and cosmic variance provides a natural lower bound on achievable uncertainties. The variance of the strain power spectrum estimator is modeled using the standard Gaussian expression, with the effective number of modes per frequency bin absorbed into an overall normalization. This approach isolates the fundamental information content of the signal and yields design-agnostic forecasts for the sensitivity of LILA to the dark energy sound speed. The methodology is identical in spirit to that employed in early sensitivity forecasts for space-based interferometers~\cite{Larson:1999we,Cutler:1997ta}, pre-design pulsar timing array cosmology studies~\cite{Jenet:2005pv,Siemens:2013zla,Sesana:2010ac}, CMB-S4 delensing projections~\cite{CMB-S4:2016ple,Abazajian:2019eic,CMB-S4:2020lpa}, and 21-cm dark ages proposals~\cite{Loeb:2003ya,Pritchard:2011xb}, where the primary goal is to establish the underlying scientific reach prior to detailed engineering optimization. More comprehensive noise modeling and mission-specific performance estimates can be straightforwardly incorporated once a concrete detector architecture is specified, but are not required to assess the qualitative or quantitative conclusions of this study.

\begin{figure}[] 
  \includegraphics[width=0.45\textwidth, keepaspectratio]{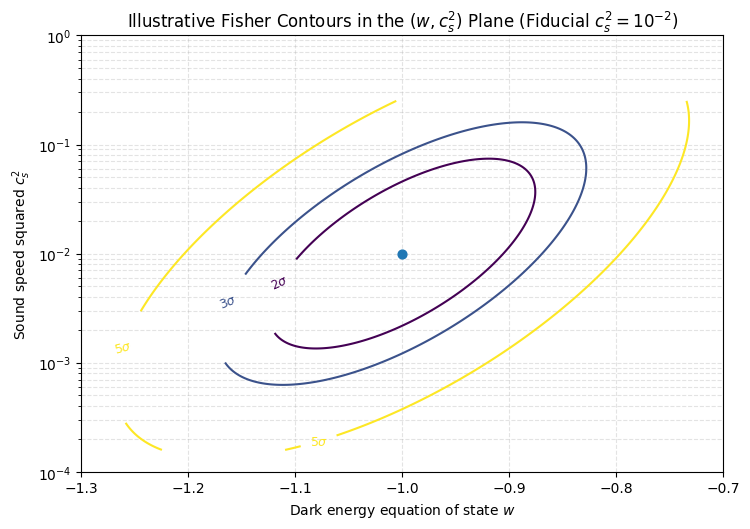}
\caption{Fisher contour ellipses in the $(w,c_s^2)$ plane for a lunar laser interferometer, centered on the fiducial clustering dark energy model $(w,c_s^2)=(-1,10^{-2})$. Shown are the joint 2$\sigma$, 3$\sigma$, and 5$\sigma$ confidence regions obtained using the cosmology-calibrated mock strain power spectrum.}
\label{fig:fisher1e-2}
\end{figure}

\begin{figure}[] 
  \includegraphics[width=0.45\textwidth, keepaspectratio]{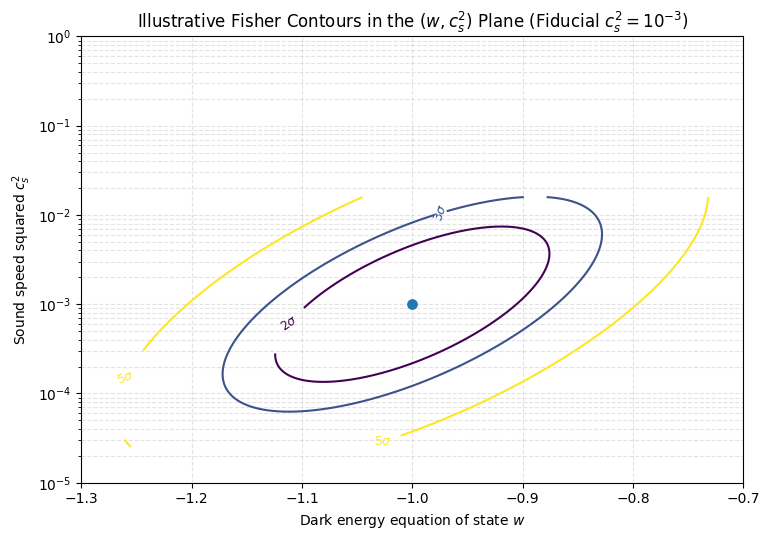}
\caption{Fisher contour ellipses in the $(w,c_s^2)$ plane for a lunar laser interferometer, centered on the fiducial clustering dark energy model $(w,c_s^2)=(-1,10^{-3})$. Shown are the joint 2$\sigma$, 3$\sigma$, and 5$\sigma$ confidence regions obtained using the cosmology-calibrated mock strain power spectrum.}
\label{fig:fisher1e-3}
\end{figure}

\section{Forecasts, Discovery, and Exclusion}

Figures~2--3 shows the illustrative Fisher contour ellipses in the $(w,c_s^2)$ plane centered on the fiducial model $(w,c_s^2)=(-1,10^{-2})$ and $(w,c_s^2)=(-1,10^{-3})$, including the 2$\sigma$, 3$\sigma$, and 5$\sigma$ joint confidence regions. 
The elongated shape of the contours reflects a partial degeneracy between the dark energy equation of state and the sound speed, arising from the fact that variations in $w$ primarily affect the background expansion history, while $c_s^2$ controls the scale-dependent clustering of dark energy through the sound horizon. The narrow extent of the contours along the $c_s^2$ direction demonstrates that the strain power spectrum measured by a lunar laser interferometer is highly sensitive to the sound speed, particularly in the ultra-low-frequency regime where dark energy perturbations contribute significantly to the gravitational potential. 

The Fisher contour analysis is updated in FIG. 3 and 4 to include a Gaussian prior constraining the dark energy equation-of-state parameter to lie within 3\% uncertainty of $w=-1$, reflecting one of the tightest widely quoted bounds from a single coherent global analysis in the final \emph{Planck} 2018 cosmological-parameters results when combined with baryon acoustic oscillations and Type~Ia supernovae~\cite{Planck2018}. Incorporating this external information significantly tightens the contours in the $w$ direction and correspondingly reduces the degeneracy-induced broadening of the sound-speed constraint, since restricting the allowed variation in $w$ limits the extent to which background expansion effects can mimic the scale-dependent spectral turnover induced by a finite dark energy sound speed. When viewed over a wider range in $w$, the contours are seen to occupy only a small region around $\Lambda$CDM, making clear that the remaining uncertainty budget is dominated by $c_s^2$. This demonstrates that the inferred constraint on the dark energy sound speed is not driven by residual freedom in the background expansion history, but instead arises from genuine sensitivity to dark energy perturbation physics encoded in the ultra-low-frequency strain spectrum.

The presence of well-separated 3$\sigma$ and 5$\sigma$ contours indicates that models with $c_s^2 \ll 1$ can be robustly distinguished from the canonical smooth dark energy case $c_s^2=1$, even after marginalizing over $w$. In particular, the exclusion of $c_s^2=1$ at high statistical significance for a fiducial clustering model provides strong evidence that LILA has discovery-level sensitivity to dark energy microphysics, enabling a qualitative test of whether dark energy clusters on cosmological scales rather than merely tightening constraints on its background equation of state.

\begin{figure}[] 
  \includegraphics[width=0.45\textwidth, keepaspectratio]{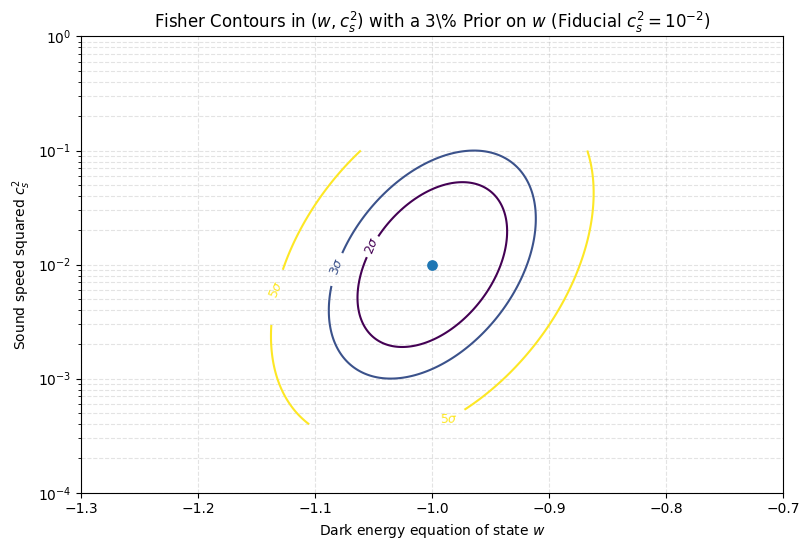}
\caption{Fisher contour ellipses in the $(w,c_s^2)$ plane for a lunar laser interferometer, centered on the fiducial clustering dark energy model $(w,c_s^2)=(-1,10^{-2})$ with $w$ constrained within 3\% of -1 using a Gaussian prior. Shown are the joint 2$\sigma$, 3$\sigma$, and 5$\sigma$ confidence regions obtained using the cosmology-calibrated mock strain power spectrum.}
\label{fig:fisher1e-2v2}
\end{figure}

\begin{figure}[] 
  \includegraphics[width=0.45\textwidth, keepaspectratio]{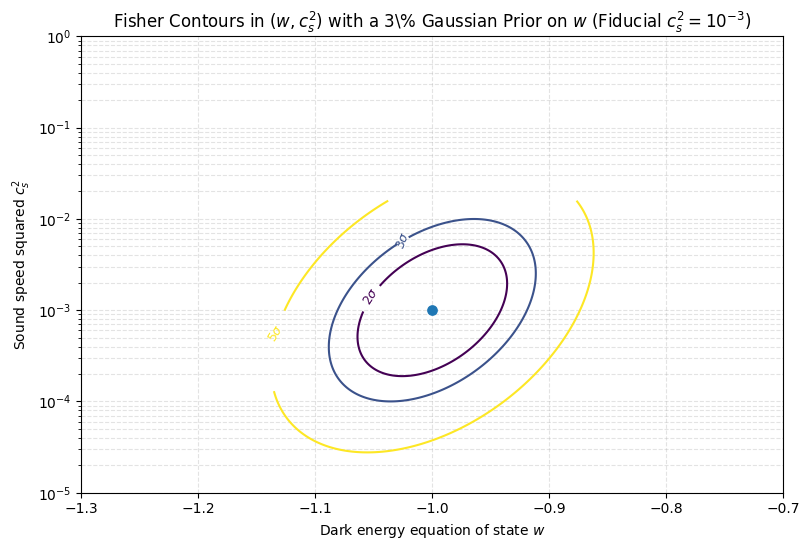}
\caption{Fisher contour ellipses in the $(w,c_s^2)$ plane for a lunar laser interferometer, centered on the fiducial clustering dark energy model $(w,c_s^2)=(-1,10^{-3})$ with $w$ constrained within 3\% of -1 using a Gaussian prior. Shown are the joint 2$\sigma$, 3$\sigma$, and 5$\sigma$ confidence regions obtained using the cosmology-calibrated mock strain power spectrum.}
\label{fig:fisher1e-3v2}
\end{figure}

This framing aligns with discovery thresholds used in high-impact cosmology experiments. Another subtlety that one should keep in mind while discussing possible sound speed constraints is how this relates to constraints on the equation of state of dark energy. Constraints on the dark energy EOS parameter are commonly parameterized in terms of $w(a)=w_0+w_a(1-a)$ or equivalent forms and they arise entirely from the background expansion history of the universe. Observables such as supernova luminosity distances, baryon acoustic oscillations and the cosmic microwave background primarily constrain the Hubble rate $H(a)$ and integrated distance measures, which depend only on the homogeneous dark energy density and pressure. At this level it should be noted that dark energy is characterized solely by its background stress-energy tensor and perturbative properties do not enter. Within the EFT framework, the background evolution is governed by the time dependent functions $\Lambda(t)$ and $c(t)$ appearing in the action and these functions determine the dark energy density and pressure through 
\begin{equation}
\rho_{\rm DE} = \Lambda + c
\end{equation}
\begin{equation}
p_{\rm DE} = -\Lambda + c
\end{equation}
leading to
\begin{equation}
w(a) = \frac{p_{\rm DE}}{\rho_{\rm DE}} = \frac{c-\Lambda}{c+\Lambda}
\end{equation}
As a result, observational constraints on $w_0$ and $w_a$ translate directly into constraints on the background combinations of $\Lambda(t)$ and $c(t)$, fixing how the dark energy density evolves with time but leaving other EFT coefficients unconstrained.

On the other hand, the squared sound speed $c_s^2$ remains a property of the perturbation sector and depends on the structure of the quadratic action for scalar fluctuations and even in the minimal EFT truncation, the sound speed is given by \eqref{basecs}, where the coefficient $M_2^4(t)$ multiplies an operator that vanishes identically on the homogeneous background. In this case, $M_2^4(t)$ has no impact on the background expansion and is therefore invisible to probes that constrain $w(a)$ and this separation persists in the more general case, where $c_s^2$ is determined by the ratio $B_{\rm eff}/A_{\rm eff}$ and depends on EFT operators that do not affect the background equations of motion. This structural separation implies that even extremely tight bounds on $w_0$ and $w_a$, including results consistent with $w=-1$ at sub-percent level, do not in themselves restrict the sound speed of dark energy. Models with identical background evolution and identical $w(a)$ can exhibit widely different clustering behavior depending on the values of perturbation-sector coefficients and in particular, scenarios with $w(a)\simeq -1$ can support sound speeds ranging from order unity to values far below unity without violating any background constraints.

The measurements discussed here therefore probe an orthogonal direction in theory space compared to traditional dark energy surveys. While background observables determine how dark energy redshifts, constraints on $c_s^2$ determine how dark energy responds to inhomogeneities and whether it clusters on horizon scales and thus, combining both types of measurements allows one to disentangle background dynamics from microphysical properties. This would allow us to break degeneracies that cannot be resolved using $w_0w_a$ constraints alone and providing a more complete characterization of the dark energy sector. 

\section{Conclusions}

We have demonstrated that a lunar laser interferometer enables a qualitatively new probe of dark energy microphysics. By measuring horizon-scale gravitational potential evolution in real time, can LILA directly constrains the sound speed of dark energy and associated EFT operators.

It is worth emphasizing that the constraints discussed here do not rely on dark energy being locally coupled to matter or remaining unscreened in high density environments. The observable targeted by a lunar laser interferometer is the time dependent evolution of scalar gravitational potentials on horizon scales which are governed by the gravitational contribution of dark energy perturbations and remains present even in theories where local fifth forces are efficiently screened. As a result, we see that the inferred constraints on the effective sound speed probe the clustering and propagation properties of dark energy at cosmological scales, and this is independent of local coupling or screening assumptions. While specific model interpretations may depend on how screening operates away from the linear regime, the core result that LILA constrains whether dark energy behaves as a smooth or clustering component, applies broadly across dark energy scenarios consistent with linear cosmological perturbation theory. 

This approach is novel, complementary to existing probes, and capable of either discovering dark energy clustering or excluding large classes of theoretical models. Our results establish lunar interferometry as a transformative tool in the study of cosmic acceleration.\\

\noindent \textbf{Acknowledgements:} We gratefully acknowledge support from Vanderbilt University and the U.S. National Science Foundation. This work is supported in part by NSF Award PHY-2411502 and by the Vanderbilt Discovery Doctoral Fellowship.

\bibliography{references}

\end{document}